# Atomic-Scale Imaging of Lattice Relaxation and Topological Flat Bands in Helical Trilayer Graphene

Shudan Jiang[1†], Zonglin Li[1†], Yu Gu[1], Liang Liu[1,4,5], Dandan Guan[1,4,5], Yaoyi Li[1,4,5], Hao Zheng[1,4,5], Canhua Liu[1,4,5], Kenji Watanabe[2], Takashi Taniguchi[3], Shengwei Jiang[1], Xiaoxue Liu[1,4,5], Zhiwen Shi[1], Guorui Chen[1], Jinfeng Jia[1,4,5,6,7*], Tingxin Li[1,4*], Can Li[1*], and Shiyong Wang[1,4,5*]

[1]State Key Laboratory of Micro-nano Engineering Science, Tsung-Dao Lee Institute & School of Physics and Astronomy, Key Laboratory of Artificial Structures and Quantum Control (Ministry of Education), Shanghai Jiao Tong University, Shanghai 200240, China

[2]Research Center for Electronic and Optical Materials, National Institute for Materials Science, 1-1 Namiki, Tsukuba 305-0044, Japan

[3]Research Center for Materials Nanoarchitectonics, National Institute for Materials Science, 1-1 Namiki, Tsukuba 305-0044, Japan

[4]Hefei National Laboratory, Hefei 230088, China

[5]Shanghai Research Center for Quantum Sciences, 99 Xiupu Road, Shanghai 201315, China

[6]Quantum Science Center of Guangdong-Hong Kong-Macao Greater Bay Area (Guangdong), Shenzhen 518045, China

[7]Department of Physics, Southern University of Science and Technology, Shenzhen 518055, China

[†]These authors contribute equally to this work.

[*]Emails: jfjia@sjtu.edu.cn, txli89@sjtu.edu.cn, lic_18@sjtu.edu.cn, shiyong.wang@sjtu.edu.cn

Abstract

Helical trilayer graphene (HTG) has emerged as a highly tunable moiré quantum material that hosts strong electronic correlations and nontrivial band topology. However, the atomic-scale lattice structure and local electronic properties have remained largely unexplored. Here we present a comprehensive real-space study of HTG using a combination of scanning near-field optical microscopy and low-temperature scanning tunneling microscopy. We directly image supermoiré lattice relaxation, revealing large triangular domains separated by sharp domain walls, as well as stripe domains connected by smoothly varying boundaries. Atomic-scale spectroscopy uncovers flat bands with a honeycomb electronic texture and one-dimensional boundary states confined to domain walls. By systematically varying the twist angle, we identify a magic angle of approximately 1.9°, substantially larger than the ~1.6° predicted by theory. Our results establish a direct microscopic link between lattice relaxation and flat bands in HTG.

The discovery of correlated insulators and unconventional superconductivity in magic-angle twisted bilayer graphene has sparked intense interest in moiré materials as a versatile platform for engineering strong electronic correlations and topological phases [1–7]. Extending beyond bilayers, twisted multilayer graphene systems provide a broader parameter space in which band topology, bandwidth, and interaction strength can be tuned through layer number, stacking chirality, and twist angle [8–31]. Among these systems, helical trilayer graphene (HTG)—formed by sequentially twisting three graphene layers with the same chirality—has emerged as a particularly promising platform. Theory predicts that HTG hosts narrow, topologically nontrivial moiré bands whose properties interpolate between those of twisted bilayer graphene and rhombohedral multilayers, with enhanced tunability via carrier density and displacement fields [32–44]. Transport and compressibility measurements have already revealed correlated insulating states and anomalous Hall phases [34,35].

A defining and distinctive feature of HTG is the real-space structure of its low-energy electronic states. The active moiré bands are predicted to form an effective honeycomb lattice, closely mimicking the Haldane model and naturally supporting topological flat bands with nonzero Chern numbers [36]. Such topological flat bands are widely regarded as promising platforms for the realization of fractional Chern insulators (FCIs), owing to the combination of narrow bandwidths, and nonzero Chern numbers [35,36,45]. Despite this progress, existing experiments on HTG, such as transport and scanning single-electron transistor measurements, reveal the presence of correlated and topological phases but necessarily average over microscopic inhomogeneities [34,35]. As a result, the atomic-scale lattice reconstruction, the real-space structure of flat-band wavefunctions, and the nature of boundary modes remain largely unexplored. This missing microscopic information leaves a critical gap in understanding how topology and interactions emerge in HTG.

In this Letter, we combine scanning near-field optical microscopy (SNOM), atomic force microscopy (AFM), and low-temperature scanning tunneling microscopy and spectroscopy (LT-STM/STS) to probe both the lattice and electronic structure of HTG at the atomic scale. SNOM and AFM directly visualize the supermoiré reconstruction and domain formation, while lattice-relaxation simulations reveal the key roles of twist-angle mismatch and strain in shaping the observed domain morphologies. LT-STM/STS reveals isolated flat bands with a honeycomb electronic texture and one-dimensional edge states confined to domain walls. By systematically studying the twist-angle dependence, we extract the flat-band bandwidth and identify a magic angle of approximately 1.9°, larger than the ~1.6° predicted by earlier theoretical work [36]. These results establish a direct microscopic link between lattice relaxation and narrow-band electronic structure in HTG, providing a foundation for future studies of correlated and topological quantum phases in this moiré platform.

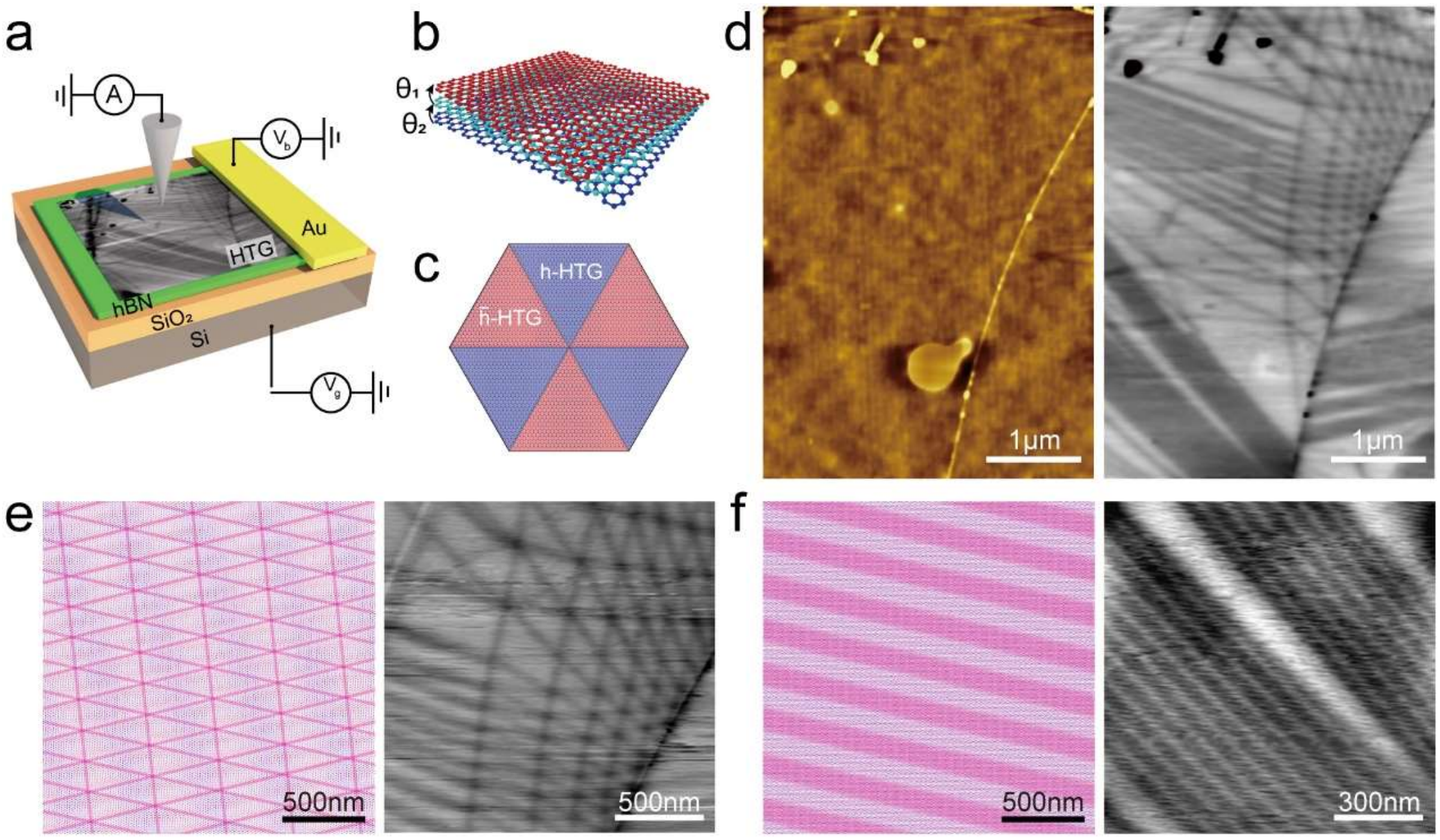

**Figure 1 | Device architecture and lattice relaxation in helical twisted trilayer graphene.** (a) Schematic of the gate-tunable device geometry used for STM/STS measurements. (b,c) Schematic illustrations of the helical twisted trilayer graphene structure and the resulting lattice reconstruction into a triangular supermoiré domain network. (d) Representative AFM topography and SNOM amplitude images of an HTG sample. (e,f) Simulated supermoiré structures (left) and corresponding zoomed-in SNOM images (right). Uniaxial heterostrain modifies the supermoiré morphology, compressing the triangular domains and driving the emergence of stripe-like domains connected by smooth boundaries, in agreement with both the relaxation simulations and SNOM observations.

As shown in Fig. 1, HTG devices were fabricated using a tear-and-stack technique that enables precise control of twist angles. Monolayer graphene exfoliated onto $SiO_2$/Si substrates was sequentially picked up with a polymer stamp and rotated by predetermined angles prior to stacking (Supplementary Fig. S1). A schematic of the device geometry is shown in Fig. 1a, where silicon or graphite gates enable electrostatic tuning of the band filling.

Figure 1b illustrates the helical twisted trilayer graphene structure and its reconstruction into a triangular supermoiré pattern. This reconstruction originates from the interference between the moiré lattices generated by the two twist angles, $\theta_1$ and $\theta_2$. For equal twist angles ($\theta_1 = \theta_2 = \theta$), the moiré patterns of the upper and lower graphene bilayers are rotated relative to one another, producing a long-wavelength supermoiré structure [36,46]. The resulting hierarchy of length scales, from the atomic lattice to the moiré and supermoiré periods, strongly amplifies the effects of lattice relaxation.

To directly visualize the reconstructed supermoiré structure, we employed SNOM, in which a metallic AFM tip illuminated by a focused infrared beam probes the local optical conductivity and stacking configuration [47]. As shown in Fig. 1d, the AFM topography is nearly featureless, indicating negligible large-scale corrugation, whereas the SNOM image reveals pronounced dark lines corresponding to supermoiré domain boundaries.

To understand the origin of these domain-wall morphologies, we performed lattice-relaxation simulations based on the model of Ref. 36. The calculations show that the supermoiré structure is highly sensitive to both twist-angle mismatch and strain. Small deviations between $\theta_1$ and $\theta_2$ significantly modify the size of the triangular domains (Supplementary Fig. S7). In addition, strain applied to one layer, or to adjacent layers with the same sign, enlarges and distorts the triangular network (Fig. 1e and Supplementary Fig. S8), whereas strains of opposite sign in adjacent layers generate stripe-like domain boundaries (Fig. 1f and Supplementary Fig. S8), consistent with the diverse domain-wall morphologies observed experimentally.

We next focus on the moiré-scale structure within a single triangular supermoiré domain. As illustrated in Fig. 2a, the AA stacking regions of the top two layers (AAB) and the bottom two layers (BAA) form the two sublattices of a periodic honeycomb lattice within each domain. This honeycomb pattern is resolved in our STM topography (Fig. 2b), which reveals two interpenetrating triangular sublattices with a well-defined periodicity of about 8.5 nm. In these images, one sublattice appears consistently brighter, a contrast we attribute to its closer vertical proximity to the STM tip and assign to the AAB stacking region.

Figure 2d compares the calculated density of states (DOS) with experimentally measured dI/dV spectra acquired at different stacking configurations. The spectra reveal a pronounced enhancement of the local DOS near charge neutrality, in good agreement with the calculated narrow-band features. Importantly, these states are well separated from the remote bands by energy gaps of approximately 100 meV, providing a favorable setting for investigating the intrinsic physics of the isolated flat-band manifold.

To further probe the filling evolution of the narrow bands, we performed gate-dependent STM/STS measurements (Fig. 2e). The low-energy Local DOS remains pinned near the Fermi level over a finite range of carrier densities and exhibits distinct features associated with fillings of approximately −4, 0, and +4 electrons per moiré unit cell. These observations provide strong evidence that the observed local DOS enhancement originates from the filling of the narrow moiré bands. Although no well-developed correlated gaps are resolved at integer fillings, likely due to the finite bandwidth and experimental energy resolution, the gate-dependent measurements establish the tunable flat-band electronic structure of HTG and provide an important foundation for future studies of interaction-driven phases.

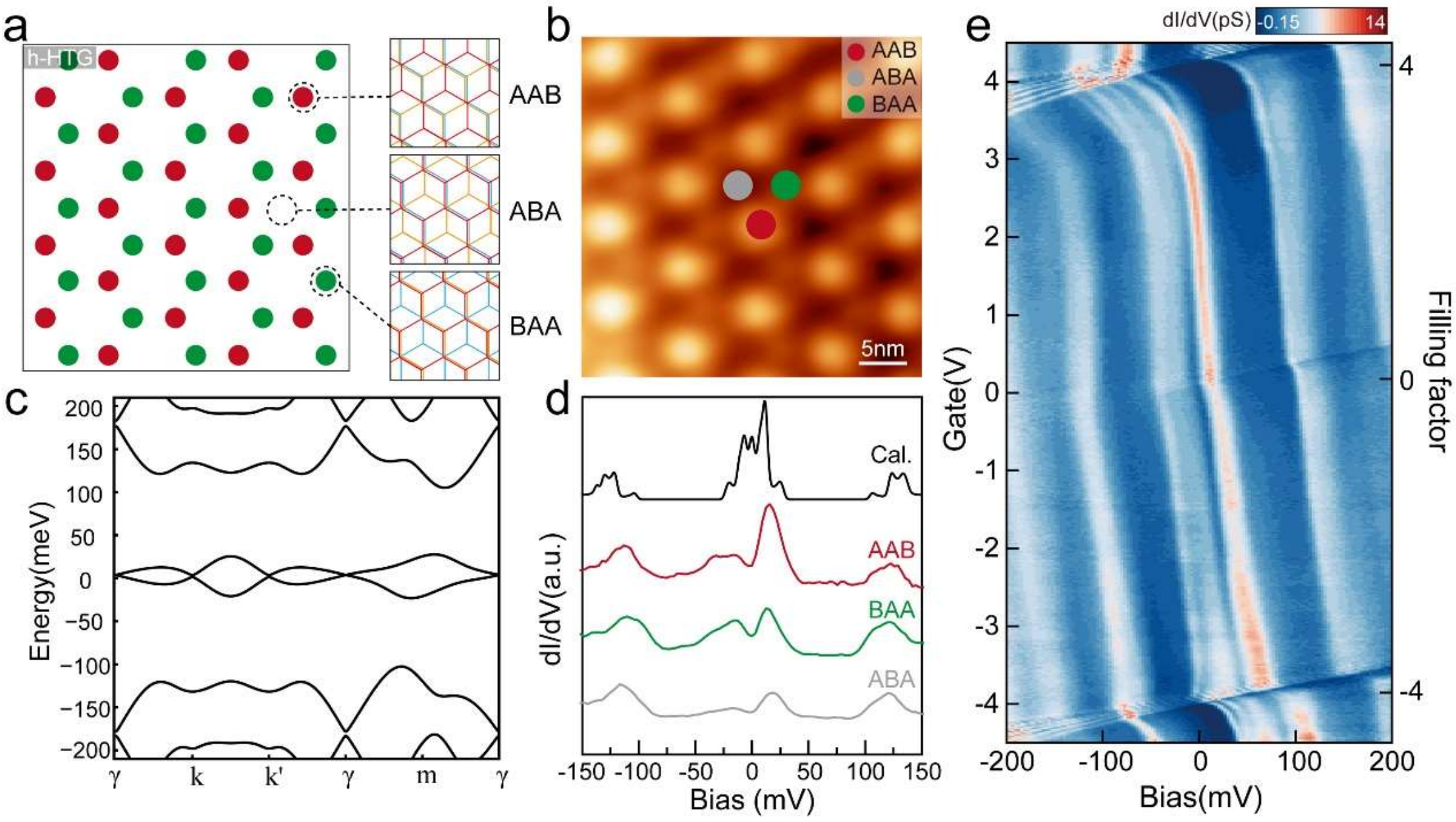

**Figure 2 | Topological flat bands and honeycomb electronic texture in HTG.** (a) Schematic illustration of the atomic stacking configurations at the high-symmetry sites of a moiré unit cell. (b) STM topography resolving the emergent honeycomb moiré lattice. (c) Calculated band structure of HTG at a twist angle of 1.65°. (d) Experimental dI/dV spectra acquired at the representative sites marked in (b), together with the calculated density of states. (e) Gate-dependent dI/dV spectra measured as a function of graphite back-gate voltage, revealing the filling evolution of the narrow moiré bands.

We now turn to the electronic structure of the domain boundaries. In HTG, the high-symmetry stacking regions correspond to AAB, ABA, and BAA configurations. Lattice relaxation produces large domains of locally periodic moiré lattices separated by sharp domain walls. Across a domain wall, the relative lateral displacement between the two moiré sublattices reverses sign, resulting in a half-moiré-lattice shift and a well-defined structural interface. We directly visualize this shift through phase-resolved FFT analysis (Supplementary Fig. S9), which reveals a characteristic π phase jump across the boundary [48,49]. Atomic-resolution STM measurements further reveal lattice relaxation associated with different local stacking configurations (Supplementary Fig. S10). Together, these observations establish the microscopic origin of the domain-wall network in HTG.

Figure 3b presents a series of dI/dV line cuts measured across a domain wall. The spectroscopic signatures of the AAB and BBA regions are nearly identical, as are those of the BAA and ABB regions, directly linking the local electronic structure to the underlying stacking configuration. This correspondence indicates that adjacent domains host inverted moiré sublattices, consistent with a reversal of the local topological character across the interface. In contrast, stripe-like domain walls exhibit a much more gradual structural and electronic evolution and do not display similarly localized boundary states (Supplementary Fig. S5).

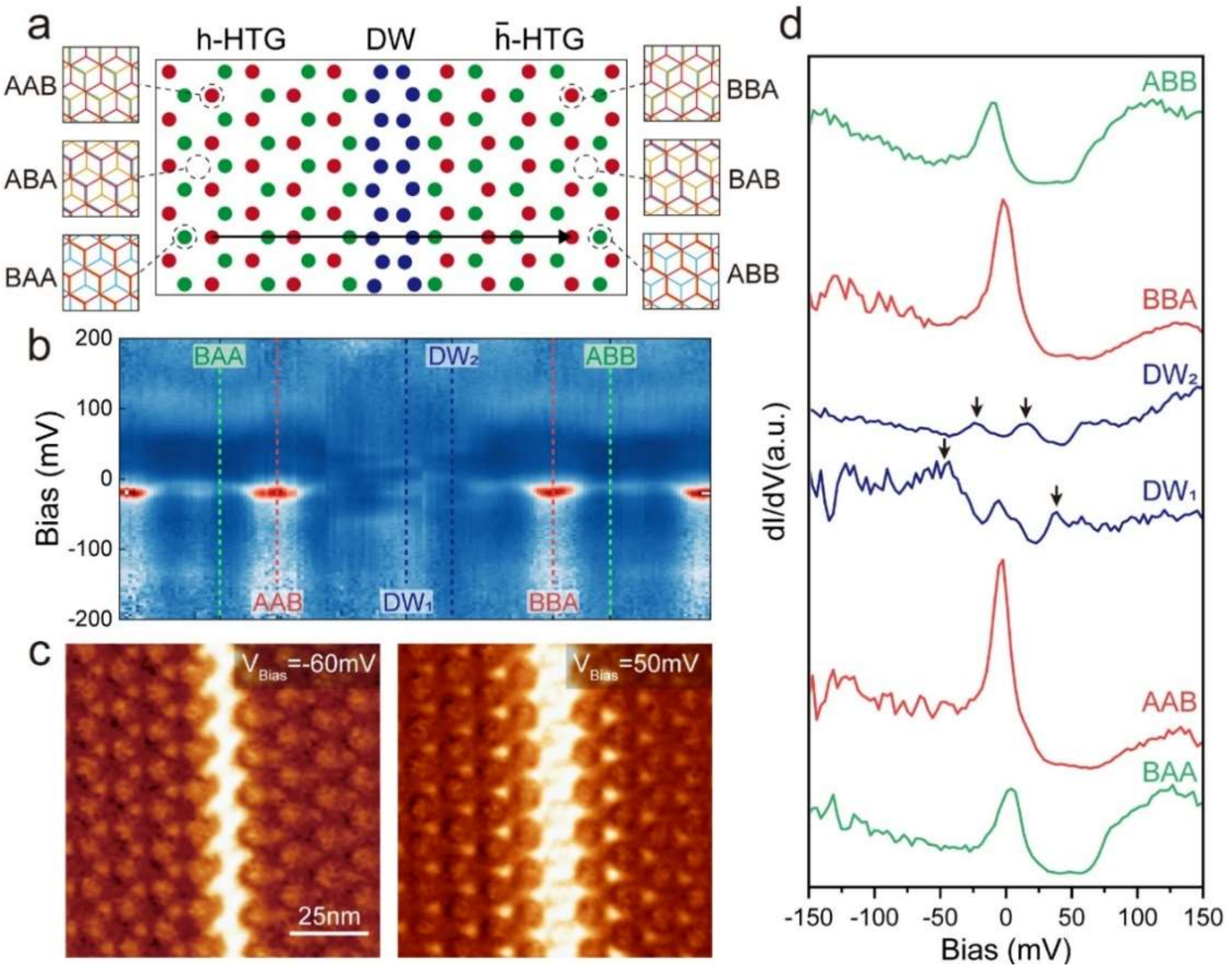

**Figure 3 | One-dimensional edge states at supermoiré domain walls.** (a) Zoomed-in schematic illustration of the atomic-scale stacking configuration across a domain wall. (b) Spatially resolved dI/dV line cut across a domain wall, revealing in-gap states. (c) dI/dV maps showing the spatial localization of one-dimensional domain-wall states. (d) Selected dI/dV spectra extracted from the positions marked in panel (b).

Beyond the bulk flat-band features, we observe a set of in-gap states localized at the domain wall, with characteristic energies near −50 meV, −25 meV, 15 meV, and 40 meV (arrows in Fig. 3d). These states are absent within the domain interiors and are strongly confined to the boundary region, as confirmed by spatially resolved dI/dV maps (Fig. 3c). The pronounced localization indicates the formation of quasi-one-dimensional electronic channels along the domain wall. Such confined states naturally emerge from the abrupt stacking transition and lattice reconstruction at the interface, which separates regions with distinct local electronic structures [47,50–53]. At present, however, our measurements do not allow us to determine the topological origin of these boundary modes or establish a direct connection to Chern-number inversion. To further investigate their nature, we performed magnetic-field-dependent STM/STS measurements up to 2.5 T (Supplementary Fig. S11). Within the accessible field range and experimental resolution, no definitive signature regarding topological Chern number was resolved.

Finally, we investigate the twist-angle dependence of the flat-band bandwidth. Figure 4a presents the calculated DOS spectra across a series of twist angles, revealing a pronounced magic-angle condition near 1.9°. At smaller angles, such as around 1.2°, the DOS shows a pair of peaks originating from the flat bands, positioned close to two higher-energy bands, resulting in four distinct spectral features near the Fermi level. As the twist angle increases, these two flat-band peaks gradually merge into a single sharp peak around 1.9°, where the bandwidth reaches a minimum.

Experimentally, dI/dV spectra measured on HTG samples with twist angles ranging from 1.2° to 2.1° (Fig. 4b) closely follow the theoretical trend. The flat-band feature becomes increasingly sharp with increasing angle and exhibits a minimum bandwidth near 1.9°, identifying the experimental magic angle. This value is notably larger than the ~1.6° predicted previously. Moreover, the increasing separation between the flat bands and remote bands with twist angle further enhances band isolation, providing a favorable setting for the emergence of interaction-driven electronic phases.

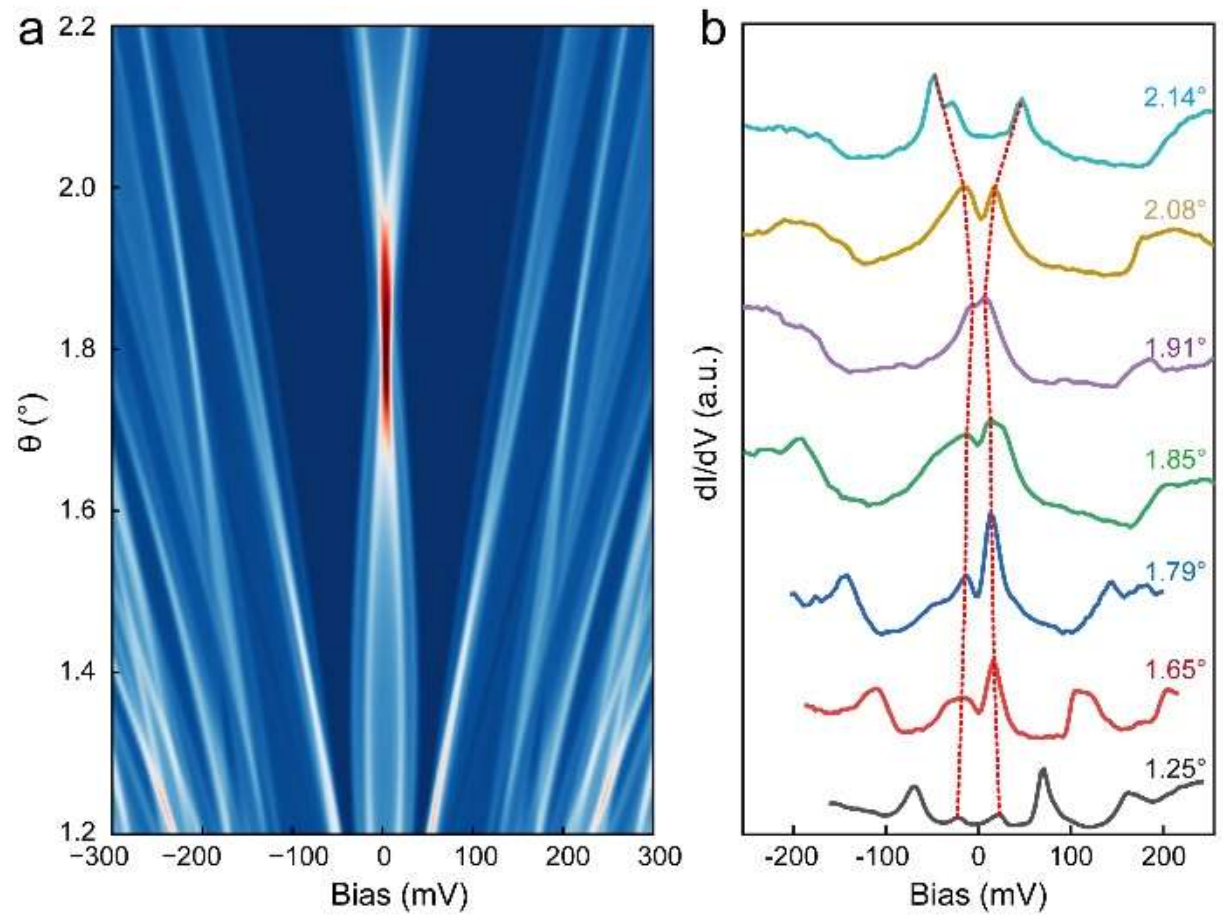

**Figure 4 | Twist-angle dependence and magic angle of helical trilayer graphene.** (a) Calculated density of states (DOS) as a function of twist angle. (b) Experimental dI/dV spectra acquired from HTG samples with different twist angles, revealing a magic angle near 1.9°.

Our combined SNOM, AFM, and STM/STS study provides a comprehensive atomic-scale characterization of helical twisted trilayer graphene. By directly correlating lattice reconstruction with the local electronic structure, we reveal the microscopic origin of its narrow bands, honeycomb electronic texture, and domain-wall states. The identification of a revised magic angle near 1.9° further highlights the important role of lattice relaxation in shaping the electronic structure of HTG. More broadly, our work establishes a real-space platform for exploring interaction-driven and topological phenomena in HTG. Future studies under lower temperatures and higher magnetic fields may enable direct visualization of fractional Chern insulators, unconventional superconductivity, and other emergent quantum phases in this highly tunable moiré system [20,39,54–56].

## Acknowledgement

We thank the NSFC (Grants No. 22325203, No. 92577203, Grants No.12488101, No. 92365302, No. 12474156, No. 12350403, No. 12474121, No. 12574146, No.12574174, No. 12304230), the Ministry of Science and Technology of China (Grants No. 2022YFA1405400, No. 2024YFA1410100, No. 2025YFE0201200, No. 2024YFF0727103), the Science and Technology Commission of Shanghai Municipality (Grants No. 24LZ1401100, No. 2019SHZDZX01, No.24LZ1401000), Cultivation Project of Shanghai Research Center for Quantum Sciences (Grant No. LZPY2024-04), Pujiang Talent Program (Grants No. 24PJA042 and No. 24PJA062) and Quantum Science and Technology-National Science and Technology Major Project (Grant No. 2021ZD0302500), and Shanghai Jiao Tong University 2030 Initiative for financial support. T. L. acknowledge support from the New Cornerstone Science Foundation through the XPLORER PRIZE.

## Author contributions

S.W., C.L., T.L. and J.J. designed and supervised the experiment. S.J., Z. L. and Y.G. fabricated the devices. S.J. and C.L. performed the STM/STS measurements. L.L., D.G., Y.L., H.Z., C.L., S.W.J., Z.S., and G.C. analyzed the data. C.L. performed theoretical studies. K.W. and T.T. grew the bulk hBN crystals. S.W. and S.J. wrote the manuscript. All authors discussed the results and commented on the manuscript.

## Data availability

The data are available from the authors upon reasonable request.